\documentclass[%
aip,
rsi,%
amsmath,
amssymb,
reprint,%
floatfix
]{revtex4-1}

\usepackage{graphicx}
\usepackage{dcolumn}
\usepackage{bm}

\usepackage{xcolor}

\begin{document}

\title[Q factor limitation at short wavelength (around 300 nm) \\in III-nitride-on-silicon photonic crystal cavities]{Q factor limitation at short wavelength (around 300 nm) \\in III-nitride-on-silicon photonic crystal cavities}

\author{Farsane Tabataba-Vakili}
\affiliation{Centre de Nanosciences et de Nanotechnologies, CNRS, Univ. Paris-Sud, Universit\'{e} Paris-Saclay, F-91405 Orsay, France.}
\affiliation{CEA, INAC-PHELIQS, Nanophysique et semiconducteurs group, F-38000 Grenoble, France.}
\author{Iannis Roland}
\affiliation{Centre de Nanosciences et de Nanotechnologies, CNRS, Univ. Paris-Sud, Universit\'{e} Paris-Saclay, F-91405 Orsay, France.}
\author{Thi-Mo Tran}
\affiliation{Centre de Nanosciences et de Nanotechnologies, CNRS, Univ. Paris-Sud, Universit\'{e} Paris-Saclay, F-91405 Orsay, France.}
\author{Xavier Checoury}
\affiliation{Centre de Nanosciences et de Nanotechnologies, CNRS, Univ. Paris-Sud, Universit\'{e} Paris-Saclay, F-91405 Orsay, France.}
\author{Moustafa El Kurdi}
\affiliation{Centre de Nanosciences et de Nanotechnologies, CNRS, Univ. Paris-Sud, Universit\'{e} Paris-Saclay, F-91405 Orsay, France.}
\author{S\'{e}bastien Sauvage}
\affiliation{Centre de Nanosciences et de Nanotechnologies, CNRS, Univ. Paris-Sud, Universit\'{e} Paris-Saclay, F-91405 Orsay, France.}
\author{Christelle Brimont}
\affiliation{Laboratoire Charles Coulomb (L2C), UMR 5221 CNRS-Universit\'{e} de Montpellier, F-34095 Montpellier, France.}
\author{Thierry Guillet}
\affiliation{Laboratoire Charles Coulomb (L2C), UMR 5221 CNRS-Universit\'{e} de Montpellier, F-34095 Montpellier, France.}
\author{St\'{e}phanie Rennesson}
\affiliation{Universit\'{e} C\^{o}te d'Azur, CRHEA-CNRS, F-06560 Valbonne, France.}
\author{Jean-Yves Duboz}
\affiliation{Universit\'{e} C\^{o}te d'Azur, CRHEA-CNRS, F-06560 Valbonne, France.}
\author{Fabrice Semond}
\affiliation{Universit\'{e} C\^{o}te d'Azur, CRHEA-CNRS, F-06560 Valbonne, France.}
\author{Bruno Gayral}
\affiliation{CEA, INAC-PHELIQS, Nanophysique et semiconducteurs group, F-38000 Grenoble, France.}
\affiliation{Univ. Grenoble Alpes, F-38000 Grenoble, France.}
\author{Philippe Boucaud}
 \email{philippe.boucaud@u-psud.fr}
 \affiliation{Centre de Nanosciences et de Nanotechnologies, CNRS, Univ. Paris-Sud, Universit\'{e} Paris-Saclay, F-91405 Orsay, France.}

\begin{abstract}
 III-nitride-on-silicon L3 photonic crystal cavities with resonances down to $315 \text{~nm}$ and quality factors (Q) up to 1085 at $337 \text{~nm}$ have been demonstrated. The reduction of quality factor with decreasing wavelength is investigated. Besides the QW absorption below $340 \text{~nm}$ a noteworthy contribution is attributed to the residual absorption present in thin AlN layers grown on silicon, as measured by spectroscopic ellipsometry. This residual absorption ultimately limits the Q factor to around 2000 at 300 nm when no active layer is present.
\end{abstract}

\maketitle

Group-III-nitride nanophotonics is a booming field with demonstrations and potential applications ranging from the near-IR to the UV-A spectral range \cite{s._arafin_notitle_2013,m._soltani_notitle_2016,a._w._bruch_notitle_2017}.
There have been several reports on III-nitride-based 1D and 2D photonic crystal (PhC) cavities in the IR \cite{n._vico_trivino_notitle_2014,w._h._p._pernice_notitle_2012,a._w._bruch_notitle_2017} and blue \cite{n._vico_trivino_notitle_2012,n._vico_trivino_notitle_2015,i._rousseau_notitle_2017,d._neel_notitle_2011, m._arita_notitle_2007} spectral ranges with large quality (Q) factors, using silicon (Si), sapphire, or silicon carbide (SiC) substrates. Fewer reports and with much lower Q factors have been made in the UV \cite{m._arita_notitle_2007, m._stegmaier_notitle_2014,d._sam-giao_notitle_2012,t.-t._wu_notitle_2013}, as the processing and the material growth are far more challenging.
Q factors around 5000 have been achieved in the blue to UV-A range at $420 \text{~nm}$ and $380 \text{~nm}$ using 2D L7 cavities \cite{n._vico_trivino_notitle_2012} and 1D nanobeam cavities \cite{s._sergent_notitle_2012,s._sergent_notitle_2012-1}, respectively. Optically pumped lasing has been achieved in L3 and H2 cavities around $370 \text{~nm}$ with Q factors up to 1700\cite{c.-h._lin_notitle_2011}. When going to shorter wavelengths $(<350\text{~nm})$ much lower Q factors ($<1000$) are observed for PhCs\cite{s._sergent_notitle_2012}. So far there have not been any good explanations for this phenomenon.

In previous work, we demonstrated Q factors of 80000 for microdisks with bus waveguides in the near-infrared \cite{i._roland_notitle_2016}. The quality factor of the microdisks decreases when going to very short wavelength  with Q factors exceeding 1000 in the range between $275 \text{~nm}$ and $470 \text{~nm}$  \cite{j._selles_notitle_2016}. The same trend was observed with photonic crystals with Q factors going from 30000-40000 at telecom wavelengths \cite{i._roland_notitle_2014,m._s._mohamed_notitle_2017} down to 4000-5000 at 380 nm \cite{d._sam-giao_notitle_2012,s._sergent_notitle_2012,s._sergent_notitle_2012-1,t.-t._wu_notitle_2013}. One open question is thus what happens at very short wavelength. 

\begin{figure}[ht]
  \centering
\includegraphics[width=1\linewidth]{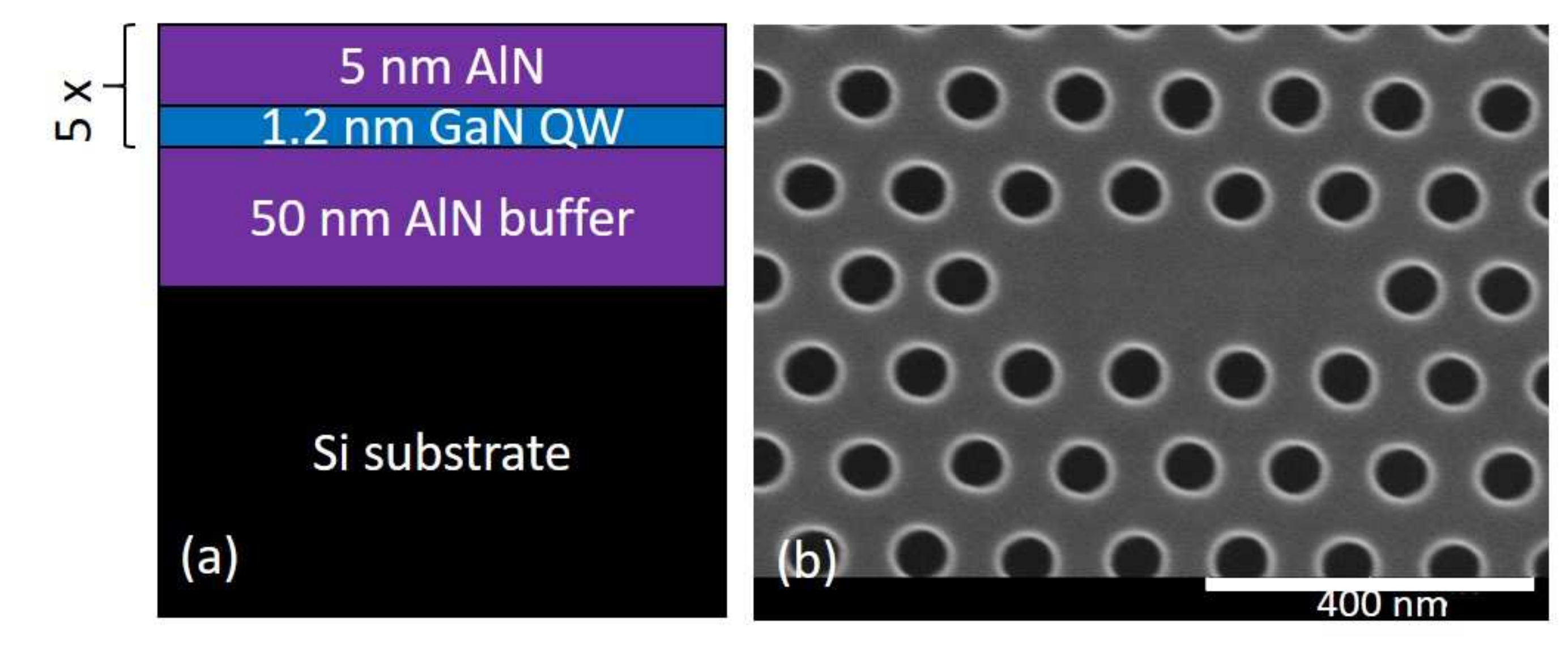}

\caption{ (a) Layer stack of the investigated sample grown on Si. (b) SEM image of a typical L3 cavity. The period $a$ is $130 \text{~nm}$, the nominal $r/a$ is 0.28, and the lateral holes of the L3 cavity are displaced by $0.10 a$. } \label{fig:1}

\end{figure}

In this work, we demonstrate resonances down to $315 \text{~nm}$ and Q factors up to 1085 at $337 \text{~nm}$ for L3 two-dimensional photonic crystal cavities. Using spectroscopic ellipsometry, we determine residual absorption in AlN grown on Si. This residual absorption is expected to be the main cause of substantially decreased Q factors at shorter wavelengths when no active layers with a large absorption (quantum wells or quantum dots) are present. 
The investigated sample for photonic crystals was grown by ammonia molecular beam epitaxy (MBE) on standard Si (111) substrate. It consists of a $50 \text{~nm}$ AlN buffer layer and 5 GaN ($1.2 \text{~nm}$)/ AlN ($5 \text{~nm}$) quantum wells (QWs) emitting at a wavelength of $310 \text{~nm}$. The heterostructure is depicted in figure \ref{fig:1}(a).

\begin{figure*}[ht]
  \centering
\includegraphics[width=0.8\linewidth]{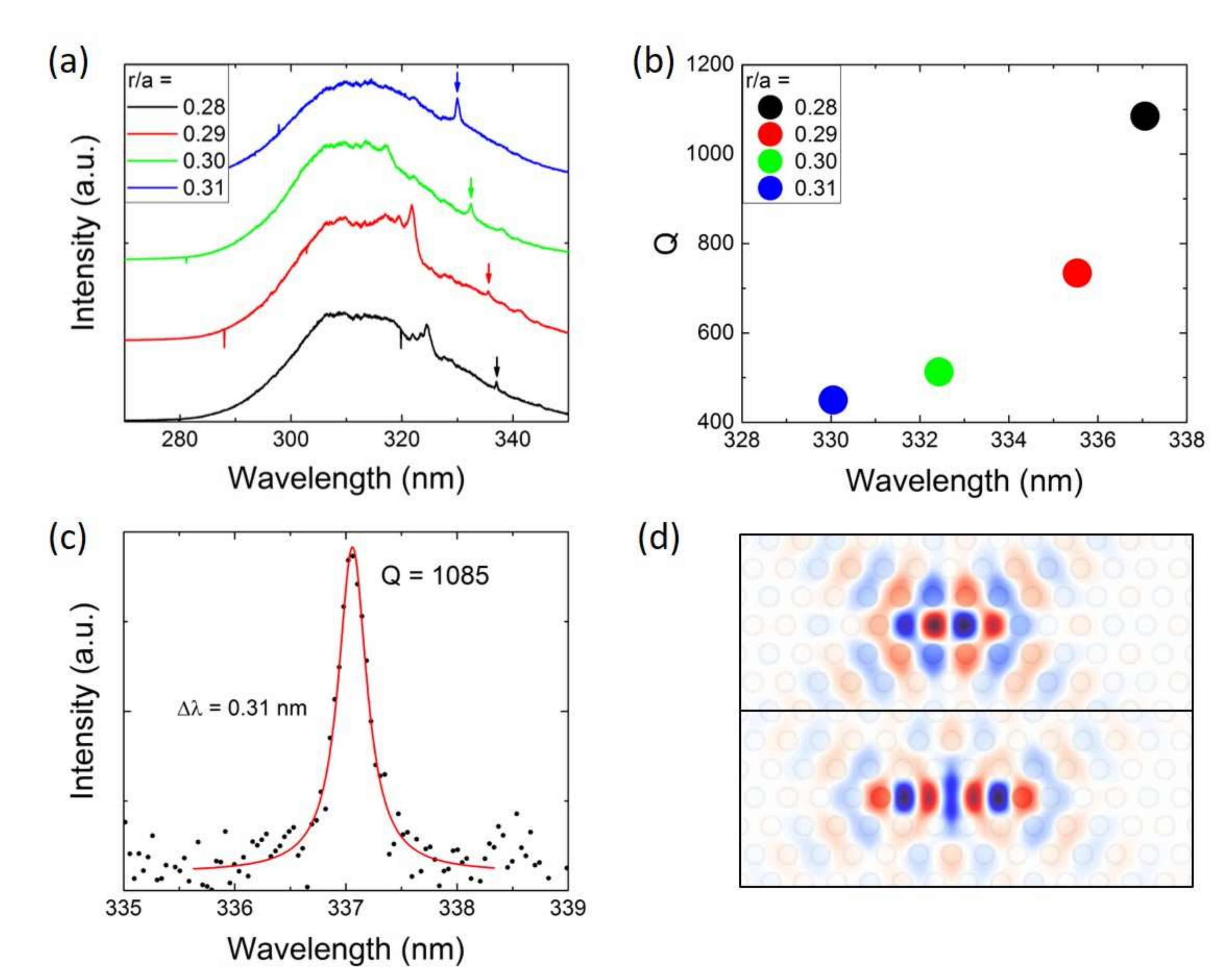}
\caption{L3 cavities with period $a= 130\text{~nm}$, hole displacement $d/a= 0.15$, and nominal radii $r/a = 0.28$ to $0.31$. (a) $\mu$PL spectra for the different $r/a $ moved along the y-axis for clarity. Arrows indicate fundamental modes. (b) Q factors for the fundamental modes in (a). (c) measured spectrum (dots) and Lorentzian fit (line) of the fundamental mode of the L3 cavity with $r/a=0.28$ shown in (a). (d) FDTD simulations of the Hz field of (top) the fundamental mode and (bottom) the first-order mode of an L3 cavity depicted in (a) with $r/a = 0.28$.} \label{fig:2}
\end{figure*}

 Standard cleanroom processing is used to fabricate triangular lattice L3 photonic crystal cavities. Electron beam lithography (EBL), reactive ion etching (RIE), and inductively coupled plasma (ICP) utilizing $\text{Cl}_2$ and $\text{BCl}_3$ gases are used. A $\text{SiO}_2$ hard mask is used to transfer the pattern from the resist into the III-N layer. Diluted ZEP resist is used for the EBL and hardened after development by electron irradiation with a scanning electron microscope (SEM) (dose $65 \text{~C/m}^2$) for higher quality RIE of the  $\text{SiO}_2$ mask. The Si substrate is then underetched using $\text{XeF}_2$ gas, resulting in suspended air hole membranes. A scanning electron microscope (SEM) image of a typical L3 cavity suspended membrane is shown in figure \ref{fig:1}(b).

We have investigated L3 cavities with periods $a = 130$ and $120\text{~nm}$ and a nominal radius over period ratio $r/a$ of $0.28$ to $0.31$. The L3 cavities correspond to the design introduced by Akahane et al.\cite{y._akahane_notitle_2003}, i.e. of  type 0, which means the two holes adjacent to the line defect are displaced\cite{e._kuramochi_notitle_2014}, in our case by $d/a = 0.15$ to $0.25$. Finite difference in time domain (FDTD) simulations show Q factors in the range of 1600 to 5000 for the L3 cavities. Larger Q factors can be achieved for smaller $r/a$. Using $d/a = 0.25$ and $r/a = 0.22$ for these L3 cavities would result in a large theoretical Q factor of more than 10000, however, such cavities are much harder to fabricate due to the much decreased hole sizes. Some theoretical papers are suggesting hole optimization schemes for L-type cavities, reaching Qs of $1-2\times 10^6$  \cite{e._kuramochi_notitle_2014, y._lai_notitle_2014}.

The PhC cavities are investigated using micro-photoluminescence ($\mu$PL) at room temperature. A frequency doubled laser emitting at $244 \text{~nm}$ is used as a pump. A microscope objective is used to focus the laser beam onto the PhC and to collect the emitted light, which is subsequently focused into the spectrometer slit with a lens and measured with a liquid nitrogen cooled charged coupled device (CCD).

Figure \ref{fig:2}(a) shows typical $\mu$PL measurements for L3 cavities with a period of $130 \text{~nm}$. The broad emission, centered around $310\text{~nm}$, comes from the QWs. The long wavelength resonances between $330$ and $340\text{~nm}$  are the fundamental modes of the L3 cavity and the shorter wavelength peaks between $320$ and $330\text{~nm}$ are the first-order modes. Shorter-wavelength modes cannot be seen because they are absorbed by the QWs. In figure \ref{fig:2}(b) the Q factors of the fundamental modes shown in figure \ref{fig:2}(a) are plotted over the wavelength. With increasing $r/a$ a blue shift and decrease in Q factor are observed, which matches well with FDTD simulations. The zoom-in of the spectrum and Lorentzian fit of the fundamental mode of the L3 cavity with $r/a = 0.28$ from figure \ref{fig:2}(a) is shown in figure \ref{fig:2}(c). The FWHM is $\Delta \lambda = 0.31 \text{~nm}$ and the Q factor is $Q=1085$ at $\lambda = 337\text{~nm}$, which is the highest we obtained. FDTD simulations of the Hz field of the fundamental (top) and first-order (bottom) modes for $r/a = 0.28$ are shown in figure \ref{fig:2}(d). An extensive classification of similar modes is reported by Chalcraft et al.\cite{a._r._a._chalcraft_notitle_2007}.

\begin{figure}[ht]
  \centering
\includegraphics[width=1\linewidth]{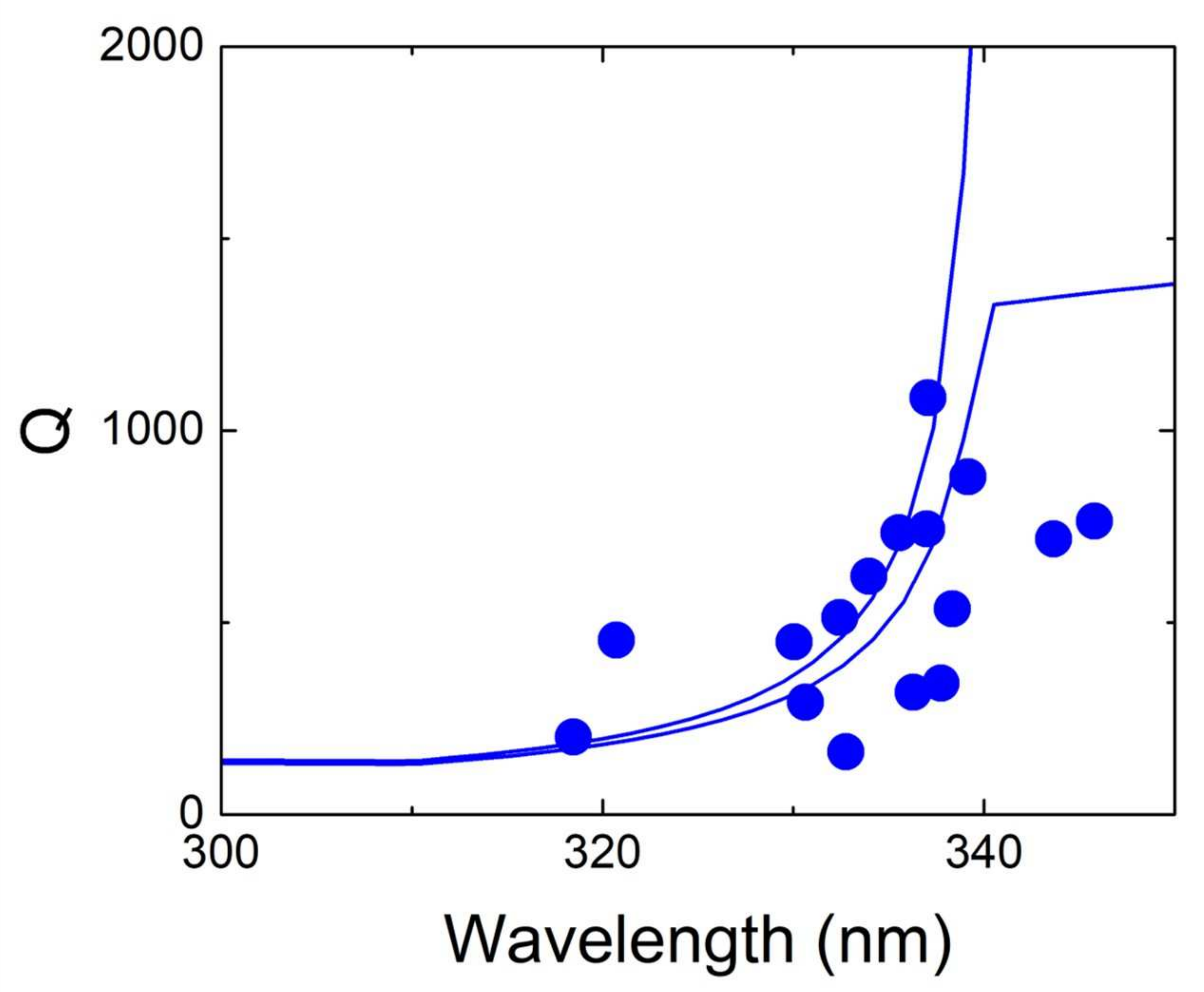}
\caption{The dots are measured Q factors of fundamental modes of various L3 cavities with varying $a$, $r/a$, and $d/a$. The lines take both the QW absorption and the residual absorption of AlN into account and are fitted using two different values of $Q_{design}$, i.e. 1600 and 5000.} \label{fig:3}
\end{figure}

Going to shorter wavelengths reduces the Q factor, as can be seen in figure \ref{fig:3} and is also observed by Sergent et al.\cite{s._sergent_notitle_2012} and Rousseau et al. \cite{i._rousseau_notitle_2017}. So far this phenomenon is rather unclear. Sidewall tapering, dispersion of holes, and surface and sidewall roughness have been suggested as the underlying cause \cite{s._sergent_notitle_2012, i._rousseau_notitle_2017}. Another factor that contributes to the reduction in Q factor is the residual absorption of AlN grown on Si. The absorption could be caused by point defects such as Al vacancies (center energy at 3.6 eV) in the AlN-Si interface region, which are known to be absorption centers in bulk AlN \cite{m._bickermann_notitle_2010,p._lu_notitle_2008,b._gil_growth_2013}. To confirm this hypothesis, we investigated several samples with varying AlN thicknesses between $50$ and $100\text{~nm}$ on Si (111) by spectroscopic ellipsometry (SE)\cite{h._fujiwara_spectroscopic_2007}. In the model we use a two-layer stack consisting of an optically unknown Cauchy film (AlN) on top of a silicon substrate for which we use literature $n$ and $k$ values, which take the substrate absorption into account. Using the Cauchy formulas we determined $n$ and $k$ for the AlN layers.

\begin{figure}[ht]
  \centering
\includegraphics[width=1\linewidth]{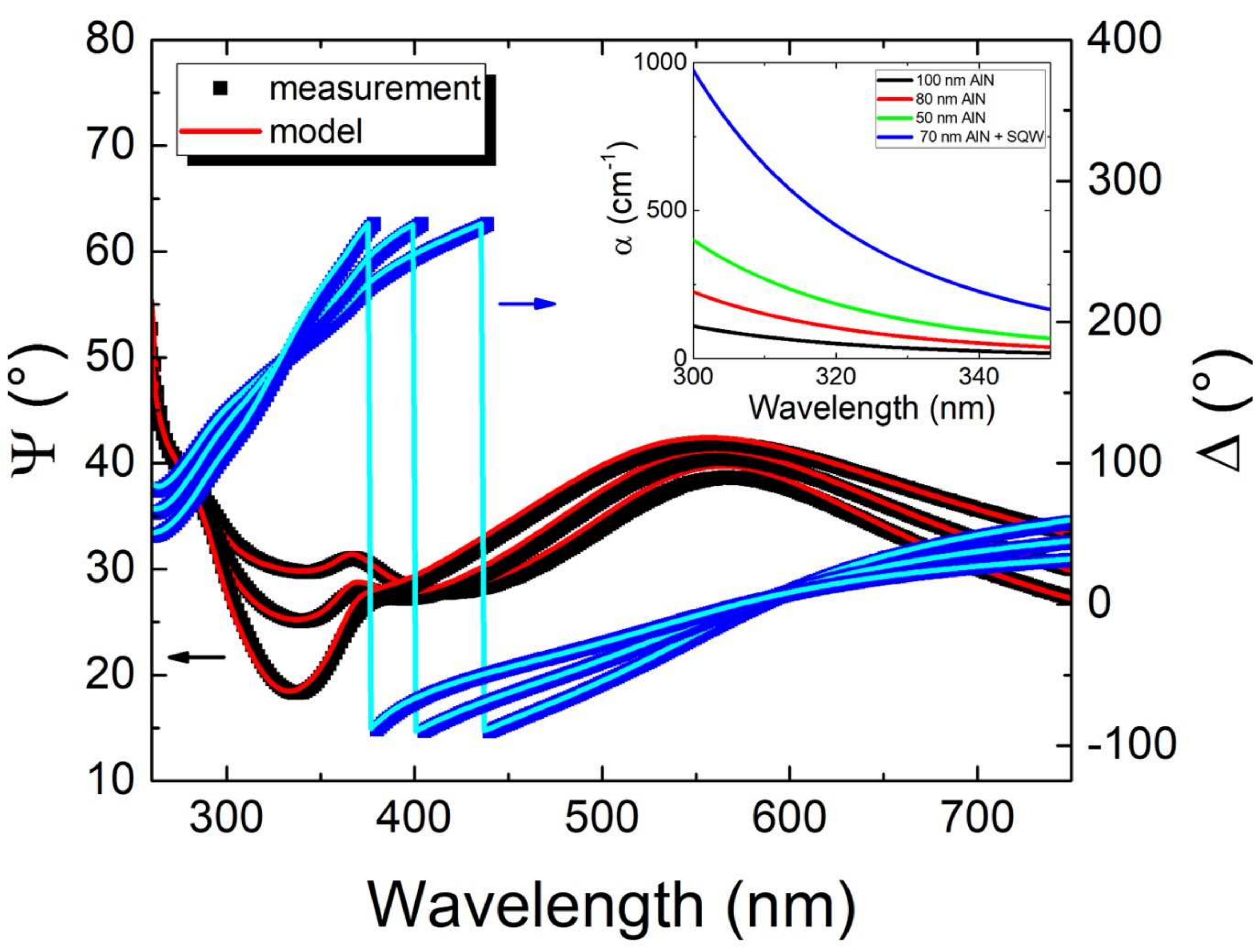}
\caption{Spectroscopic ellipsometry measurement (squares) and model (lines) for angles of incidence of $65^\circ$, $70^\circ$ and $75^\circ$ for a sample with $80 \text{~nm}$ of AlN grown on Si (111). The inset shows the AlN absorption coefficient $\alpha$ for samples with different AlN thicknesses.} \label{fig:4}
\end{figure}

\begin{eqnarray}
n & = & A + \frac{B}{\lambda ^2}+ \frac{C}{\lambda^4},\\
k & =& \alpha \exp\beta(12400(\frac{1}{\lambda}-\frac{1}{\gamma})), 
\end{eqnarray}

with $A$, $B$, $C$, $\alpha$, $\beta$ being fit parameters\cite{j._m._khoshman_notitle_2005}. The fits were performed by taking $n$ from the Cauchy model of the sample with 50 nm AlN and determining $k$ by fitting the ellipsometry data of each sample to the Cauchy model ($n$ and $\beta$ constant for all samples).
Figure \ref{fig:4} shows SE measurements (squares), highlighting the standard phase difference $\Psi$ and the amplitude ratio $\Delta$, and a Cauchy fit model (lines) for a sample with $80 \text{~nm}$ AlN for angles of incidence of $65^\circ$, $70^\circ$ and $75^\circ$. The set of Cauchy parameters that gives a good fit of $\Psi$ and $\Delta$ is not unique, resulting in small variations of $n$ and $k$. The inset in figure \ref{fig:4} shows the absorption coefficient $\alpha$ for the samples with different AlN thickness and a sample with a thin QW, where the AlN buffer layer was grown under different conditions (lower ammonia). The absorption coefficient $\alpha$ was determined using

\begin{equation}
\alpha_{AlN} = \frac{4 \pi k}{\lambda}.\label{eq:alpha}
\end{equation}

The trend first shows a systematic increase of absorption when going to shorter wavelengths. It also shows an increase in absorption with a decrease in thickness. However, the sample with the thin QW and the lower ammonia shows a higher absorption, although the QW absorption is outside of the depicted range at around $280 \text{~nm}$. Presumably the material absorption is strongly dependent on the growth parameters and ammonia flux.

The full lines in figure \ref{fig:3} show the fit model of the total Q factor of the PhC sample determined by

\begin{equation}
\frac{1}{Q_{tot}} = \frac{1}{Q_{abs,QW}} +\frac{1}{Q_{abs,AlN} }+\frac{1}{Q_{design}}\label{eq:Qtotal},
\end{equation}

where $Q_{abs,AlN}$ is given by

\begin{equation}
Q_{abs,AlN} =\frac{2 \pi n}{\alpha \lambda},\label{eq:Q}
\end{equation}

and is determined using $\alpha$ from the 80 nm AlN sample. 

The 5 QWs contribute to the absorption that limits the Q factor. We approximate the QW absorption as $Q_{abs,QW} = 3000\text{~cm}^{-1}$ below $310 \text{~nm}$ using

\begin{equation}
\alpha_{QW} \times L \approx \frac{\pi e^2}{2 h}\times \Gamma = 6\times 10^{-3} \times\Gamma,
\end{equation}

with the thickness of the QW $L=1.2\text{~nm}$, the overlap factor between the 5 QWs and the mode $\Gamma\approx 0.06$, the fundamental charge $e$, and Planck's constant $h$\cite{j._h._davies_physics_1998}. We assume a linear decrease of $\alpha$ down to $0\text{~cm}^{-1}$ at $340\text{~nm}$, close to the PL cut-off. We consider two values of $Q_{design}$, 1600 and 5000 from FDTD simulations, giving two results for the total Q, as shown in figure \ref{fig:3}. The effect of $Q_{design}$ is minimal in this range. The modeling matches very well with the measured data. This means that for the here investigated PhC sample the limitation in Q factor stems mainly f
rom QW absorption and partly from residual absorption in AlN. However, considering samples with fewer QWs or quantum dots (QDs), where the absorption by the active region would be significantly reduced compared to here, the residual absorption in AlN grown on Si would ultimately be the limitation for the Q factor at short wavelengths, i.e. 2000 at 300 nm, the latter value being obtained from the measured absorption of the 80 nm thick sample. 

The results in the inset of figure \ref{fig:4} indicate a reduction in absorption with increasing AlN thickness, which may be related to the material quality improving during the longer growth. Moreover in a thicker AlN layer the mode sees less of the more absorptive  interface region. It explains why higher Q factors could be obtained in thicker microdisks at the same wavelength.   

In conclusion, we have demonstrated high Q factor resonances in the UV-A with L3 photonic crystal cavities fabricated on a III-nitride-on-silicon sample. The Q factor at $340 \text{~nm}$ is limited experimentally to around 1000.  Residual absorption in AlN grown on Si is a significant limiting factor when no active layers are present. Our results indicate that achieving high Q factors at short wavelengths down to the UV-C is extremely challenging for very thin AlN layers ($< 100 \text{~nm}$) grown on silicon and would require specific optimization of the material growth.

The authors acknowledge support from the projects GANEX (ANR-11-LABX-0014) and QUANONIC (ANR- 13-BS10-0010). GANEX belongs to the public funded ''Investissements d'Avenir'' program managed by the French ANR agency. This work was partly supported by the RENATECH network. This work is also partially supported by a public grant overseen by the French National Research Agency (ANR) as part of the 'Investissements d'Avenir program' (Labex NanoSaclay, reference: ANR-10-LABX-0035).

\bibliography{bib.bib}

\end{document}